\documentclass[aps,preprintnumbers,showpacs, nofootinbibt,referee,keywords,twocolumn]{revtex4}
\usepackage{graphicx}
\usepackage{amsmath}
\usepackage{bm}
\usepackage{color}

\def\be{\begin{equation}}
\def\ee{\end{equation}}
\def\bea{\begin{eqnarray}}
\def\eea{\end{eqnarray}}

\begin{document}

\title{Dark matter density profile and galactic metric in Eddington-inspired
Born-Infeld gravity}
\author{Tiberiu Harko$^1$}
\email{t.harko@ucl.ac.uk}
\author{Francisco S.~N.~Lobo$^{2}$}
\email{flobo@cii.fc.ul.pt}
\author{M. K. Mak$^{3}$}
\email{mkmak@vtc.edu.hk}
\author{Sergey V. Sushkov$^{4}$}
\email{sergey_sushkov@mail.ru}
\affiliation{$^1$Department of Mathematics, University College London, Gower Street,
London WC1E 6BT, United Kingdom}
\affiliation{$^2$Centro de Astronomia e Astrof\'{\i}sica da Universidade de Lisboa, Campo
Grande, Ed. C8 1749-016 Lisboa, Portugal}
\affiliation{$^{3}$Department of Computing and Information Management, Hong Kong
Institute of Vocational Education, Chai Wan, Hong Kong,}
\affiliation{$^4$Institute of Physics, Kazan Federal University, Kremlevskaya Street 18,
Kazan 420008, Russia}
\date{\today}

\begin{abstract}
We consider the density profile of pressureless dark matter in
Eddington-inspired Born-Infeld (EiBI) gravity. The gravitational field
equations are investigated for a spherically symmetric dark matter galactic
halo, by adopting a phenomenological tangential velocity profile for test
particles moving in stable circular orbits around the galactic center. The
density profile and the mass distribution, as well as the general form of the metric tensor is obtained by numerically integrating the gravitational field equations, and in an approximate
analytical form  by using the Newtonian limit of the theory. In the weak field limit the dark matter density distribution is described by the Lane-Emden equation with polytropic index $n=1$, and is non-singular at
the galactic center. The parameter $\kappa $ of the theory is determined so that that the theory could provide a realistic description of the dark matter halos. The gravitational properties of the dark matter halos
are also briefly discussed in the Newtonian approximation.
\pacs{04.50.Kd,04.20.Cv}
\end{abstract}

\maketitle




\section{Introduction.}

Despite numerous observational and experimental
attempts, up to now the elusive dark matter particles have evaded detection.
Great hopes in gaining more insights into the nature of dark matter are
related to the Alpha Magnetic Spectrometer (AMS-02) collaboration results,
which have just been released \cite{AMS}. The AMS experiment has determined
the positron fraction in cosmic rays with extremely high precision. The
positron fraction rises continuously from $\sim 5 $ GeV up to $\sim 350$
GeV, while the slope becomes flat above $\sim 100$ GeV \cite{AMS}. In order
to explain the AMS results, dark matter is still an attractive candidate. In
fact, dark matter annihilation into $\tau ^{+}\tau ^{-}$ final states, which
results in a soft positron spectrum can account for the AMS-02 data quite
well. Other channels for dark matter annihilation/decay into multiple $\mu $
or $\tau $ leptons can also reproduce the AMS-02 data \cite{DM}. However,
presently, the only \textit{convincing} evidence, based on the flat rotation
galactic curves and the virial mass discrepancy in clusters of galaxies for
the existence of dark matter, is \textit{gravitational} \cite{3a}-\cite{3e}.

In standard general relativity the coupling between matter and gravity is
given by a proportionality relation between the stress-energy tensor and the
geometry. Although both the stress-energy tensor and the Einstein tensor are
divergenceless, there is no obvious reason why the matter-gravity coupling
should be linear \cite{4a,4b}. On the other hand, modified
theories of gravity usually affect the vacuum dynamics, yet keep the
matter-gravity coupling linear. Recently, to provide matter-gravity coupling
modifications, based on the Eddington gravity action \cite{4}, and
Born-Infeld nonlinear electrodynamics \cite{5}, the Eddington-inspired
Born-Infeld (EiBI) theory has been proposed \cite{0,1}. The EiBI theory
coupled to a perfect fluid reduces to standard general relativity coupled to
a nonlinearly modified perfect fluid, leads to an ambiguity between modified
coupling and a modified equation of state \cite{2,Pani}. The theory has
interesting cosmological consequences, leading to the possibility of
avoiding cosmological singularities \cite{cosm}. The structure of the neutron and quark stars for different equations of state of the dense matter were investigated in \cite{new}. However, the EiBI theory is
reminiscent of the Palatini gravity formulation, and it shares the same
pathologies, such as curvature singularities at the surface of polytropic
stars \cite{3}.

It is the purpose of this Letter to investigate the properties of dark matter
halos in the EiBI gravity theory. By assuming a spherically symmetric pressureless
dark matter halo, and a very general tangential rotation curve profile, the gravitational field equations of the EiBI model are exactly solved numerically, and the behavior of the dimensionless density and mass profiles, as well as the metric tensor coefficients are explicitly obtained. The dark matter distribution has a very sharp boundary that defines the radius of the dark matter distribution. In order to obtain an analytic representation of the dark matter halo properties the gravitational field equations describing the dark matter distribution are
solved to a first order approximation of the coupling constant $\kappa $.
The dark matter density profile and the galactic metric in the dark matter
halo is explored, and explicitly obtained, and the respective gravitational properties are also
briefly discussed in the Newtonian approximation.

The present paper is organized as follows. The Eddington-Inspired Born-Infeld gravity is briefly reviewed in Section~\ref{sect2}. The basic field equations describing the galactic dark matter halo properties are written down in Section~\ref{sect3}.  In Section~\ref{sect4} we  solve numerically the field equations describing dark matter halos, and obtain the density and the mass distributions, as well as the metric coefficients. An approximate solution of the field equations is  obtained, and its properties are discussed in Section~\ref{new}.   We discuss and conclude our results in Section~\ref{sect5}.

\section{Eddington-Inspired Born-Infeld gravity}\label{sect2}

 The starting point of the
EiBI theory is the gravitational action $S$, given by \cite{1},
\begin{eqnarray}
S &=&\frac{1}{8\pi}\frac{1}{\kappa }\int d^{4}x\left( \sqrt{-\left| g_{\mu
\nu }+\kappa R_{\mu \nu }\right| }-\lambda \sqrt{-g}\right) + \notag \\
&&S_{M}\left[ g,\Psi _{M}\right] ,  \label{1c}
\end{eqnarray}%
where $\lambda \neq 0$ is a dimensionless parameter, and $\kappa $ is a
parameter with inverse dimension to that of the cosmological constant $%
\Lambda $. $R_{\mu \nu }$ is the symmetric part of the Ricci tensor, and is
constructed solely from the connection $\Gamma _{\beta \gamma }^{\alpha }$.
The matter action $S_{M}$ depends only on the metric $g_{\mu \nu }$ and the
matter fields $\Psi _{M}$. The determinant of the tensor $g_{\mu \nu }+
\kappa R_{\mu \nu }$ is denoted by $\left| g_{\mu \nu }+\kappa R_{\mu
\nu }\right| $. In the limit $\kappa \rightarrow 0$, the action Eq.~(\ref{1c}%
) recovers the Einstein-Hilbert action with $\lambda = \Lambda \kappa +1$%
. However, in the present paper, we consider only asymptotic flat solutions,
and hence take $\lambda =1$. Therefore the cosmological constant vanishes,
and the remaining parameter $\kappa $ plays the determining role for the
description of the physical behavior of various cosmological and stellar
scenarios. Several constraints on the value and the sign of the parameter $%
\kappa $ have been obtained from solar observations, Big Bang
nucleosynthesis, and the existence of neutron stars in \cite{1,7,8na,8nb,8nc}. The
structure of compact stars in EiBI theory has been investigated by several
authors \cite{7,10a,10b}. In particular, for cases with positive $\kappa $,
effective gravitational repulsion prevails, leading to the existence of
pressureless stars (stars made of non-interacting particles which provide
interesting models for self-gravitating dark matter) and increases in the
mass limits of compact stars \cite{7,8na,8nb,8nc, 10a,10b}.

In the EiBI theory the metric $g_{\mu \nu }$ and the connection $\Gamma
_{\beta \gamma }^{\alpha }$ are treated as independent fields. By varying
the action (\ref{1c}) with respect to the connection $\Gamma _{\beta \gamma
}^{\alpha }$, and with respect to the real metric $g_{\mu \nu }$, yield the
following field equations 
$q_{\mu \nu }=g_{\mu \nu }+ \kappa R_{\mu \nu }$ and $q^{\mu \nu }=\tau
\left( g^{\mu \nu }-8\pi \kappa T^{\mu \nu }\right)$, 
respectively. The auxiliary metric $q_{\mu \nu }$ is related to the
connection by $\Gamma _{\beta \gamma }^{\alpha }=\frac{1}{2}q^{\alpha \sigma
}\left( \partial _{\gamma }q_{\sigma \beta }+\partial _{\beta }q_{\sigma
\gamma }-\partial _{\sigma }q_{\beta \gamma }\right) $, and $\tau$ is
defined as $\tau =\sqrt{g/q}$. If the stress-energy tensor $T^{\mu \nu }$
vanishes, then the real metric $g_{\mu \nu }$ is equal to the apparent
metric $q_{\mu \nu }$. Hence, in vacuum the EiBI theory is equivalent to standard general relativity.
The stress-energy tensor $T^{\mu \nu }$ satisfies the conservation equations
given by $\nabla _{\mu }T^{\mu \nu }=0$, where the covariant derivative $%
\nabla _{\mu }$ refers to the metric $g_{\mu \nu }$.

\section{Static spherically symmetric dark matter halos in EiBI gravity}\label{sect3}

We consider that the dark matter halo is static and spherically symmetric.
Therefore the line elements for the physical metric $g_{\mu \nu }$ and for
the auxiliary metric $q_{\mu \nu }$ take the forms
\begin{eqnarray}
g_{\mu \nu }dx^{\mu }dx^{\nu }&=&-e^{\nu \left( r\right) }c^2dt^{2}+e^{\lambda
\left( r\right) }dr^{2}+f\left( r\right) d\Omega ^{2},  \label{8c} \\
q_{\mu \nu }dx^{\mu }dx^{\nu }&=&-e^{\beta \left( r\right) }c^2dt^{2}+e^{\alpha
\left( r\right) }dr^{2}+r^{2}d\Omega ^{2},  \label{9c}
\end{eqnarray}
respectively, where $\nu \left( r\right) $, $\lambda \left( r\right) $, $\beta \left(
r\right) $, $\alpha \left( r\right) $ and $f\left( r\right) $ are arbitrary
metric functions of the radial coordinate $r$, and $d\Omega ^{2}=d\theta
^{2}+\sin ^{2}\theta d\phi ^{2}$. The galactic halo is made up of a perfect
fluid described by the standard stress-energy tensor
\be
T^{\mu \nu }=\left(
\rho c^2+p\right) u^{\mu }u^{\nu }+pg^{\mu \nu },
\ee
where $\rho $, $p$ and $%
u^{\mu }$ are the energy density, the isotropic pressure and the four
velocity of the fluid, respectively, with the latter satisfying $u^{\mu
}u^{\nu }g_{\mu \nu }=-1$. The system of gravitational field equations
describing the dark matter halo structure yields \cite{10a,10b}
\begin{eqnarray}
\frac{d}{dr}\left(re^{-\alpha }\right)=1-\frac{1}{2\kappa }\left( 2+\frac{a}{%
b^{3}}-\frac{3}{ab}\right)r^2 ,  \label{g1} \\
e^{-\alpha }\left(1+r\frac{d\beta }{dr}\right)=1+\frac{1}{2\kappa }\left(
\frac{1}{ab}+\frac{a}{b^{3}}-2\right)r^2,  \label{g2} \\
e^{\beta }=\frac{e^{\nu }b^{3}}{a}, \qquad e^{\alpha }=e^{\lambda }ab,
\qquad f=\frac{r^{2}}{ab},
\end{eqnarray}
where we have defined the arbitrary functions $a\left( r\right) $ and $%
b\left( r\right) $ as
\begin{equation}
a=\sqrt{1+8\pi \kappa \frac{G}{c^2} \rho },
\ee
and
\be
b=\sqrt{1-8\pi \kappa \frac{G}{c^4} p},
\end{equation}
respectively. In the $g$--metric the conservation of the stress-energy
tensor yields the result
\begin{equation}
\frac{d\nu }{dr}=-\frac{2}{p+\rho c^2}\frac{dp}{dr}=\frac{4b}{a^{2}-b^{2}}\frac{%
db}{dr}.  \label{psi}
\end{equation}

The existence of dark matter at the galactic scale is inferred from the
study of the rotational velocities of massive test particles (hydrogen
clouds) around the galactic center. The Lagrangian $\mathcal{L}$ for a
massive test particle reads
\begin{equation}
\mathcal{L}=\frac{1}{2}( -e^{\nu }c^2\dot{t}^{2}+e^{\lambda }\dot{r}^{2}+r^{2}
\dot{\Omega}^{2}), \label{lag}
\end{equation}
where the overdot denotes differentiation with respect to the affine
parameter $s$. By defining the tangential velocity $v_{\mathrm{tg}}$ of a
test particle, as measured in terms of the proper time~\cite{LaLi}, that is,
by an observer located at a given point, as $v_{tg}^{2}=e^{-\nu }r^{2}\left(
d\Omega /dt\right)^2$, we obtain the expression of $v_{tg}^{2}$ for a test
particle in a stable circular orbit as ~\cite{Matosa} - \cite{Matose}
\be
\frac{v_{tg}^{2}}{c^2}=\frac{1}{2}r\nu ^{\prime }.
\ee
For $v_{tg}^{2}$ we assume the simple empirical dark halo rotational
velocity law \cite{per}
\begin{equation}
v_{tg}^{2}=v_{0\infty}^{2}\frac{\left(r/r_{opt}\right)^{2}}{%
\left(r/r_{opt}\right)^{2}+r_0^2},  \label{vel}
\end{equation}%
where $r_{opt}$ is the optical radius containing 83$\%$ of the galactic
luminosity. The parameter $r_0$, defined as the ratio of the halo core
radius and $r_{opt}$, and the terminal velocity $v_{\infty }$ are functions
of the galactic luminosity $L$. For spiral galaxies $r_0=1.5\left( L/L_{\ast
}\right) ^{1/5}$ and $v_{\infty}^{2}=v_{opt}^{2}\left( 1-\beta _{\ast
}\right) \left( 1+r_0^{2}\right) $, where $v_{opt}=v_{tg}\left(
r_{opt}\right) $, and $\beta _{\ast }=0.72+0.44\log _{10}\left( L/L_{\ast
}\right) $, with $L_{\ast }=10^{10.4}L_{\odot}$. For $r\rightarrow 0 $, $%
v_{tg}\rightarrow 0$, while for $r/r_{opt} \gg r_0$, $v_{tg}\rightarrow
v_{0\infty}$. In the following we denote $v_{\infty}$ as
\be
v_{\infty}^2=\frac{v_{0\infty}^2}{c^2}.
\ee

Therefore, the most general static and spherically symmetric metric of the
dark matter halo can be written as
\begin{equation}
ds^{2}=-e^{\nu _0}\left[ \left(\frac{r}{r_{opt}}\right)^2+r_0^2\right]
^{v_{\infty}^{2}}dt^{2}+e^{\lambda (r)}dr^{2}+r^{2}d\Omega ^{2},  \label{nu1}
\end{equation}
where $e^{\nu _0}$ is an arbitrary constant of integration.

In the Newtonian limit the $%
g_{tt}$ component of the metric tensor is given by $e^{\nu}\approx
1+2\Phi_{N}$, where $\Phi _{N}$ is the Newtonian gravitational potential
satisfying the Poisson equation $\Delta \Phi_{N}=4\pi\rho$~\cite{LaLi}. In
the constant velocity region the mass $M(r)$ of the dark matter and the
energy density $\rho$ vary with the distance as $M(r)=v_{\mathrm{tg}}^{2}r$
and $\rho =v_{\mathrm{tg}}^{2}/4\pi r^{2}$, respectively.

\section{The dark matter density profile and the galactic metric in EiBI gravity}\label{sect4}

In the present Section we will restore the normal astrophysics units in all equations. In order to study the extra-galactic motion of massive test
particles in EiBI gravity, we assume that the motion takes place in a
 cosmic medium that satisfies the $\left(8\pi G\kappa/c^4\right) p\ll 1$. Therefore  we immediately obtain $%
b\approx 1$, and hence the dark matter properties can be described by the
effective matter density $\rho $ only, which determines the function $a$.  For a galactic metric of the form given by Eq.~(\ref{nu1}) we
obtain
\begin{eqnarray}
e^{\beta } &=&\frac{e^{\nu _{0}}\left[ \left( r/r_{opt}\right) ^{2}+r_{0}^{2}%
\right] ^{v_{\infty }^{2}}}{\sqrt{1+8\pi \kappa \rho }}, \\
\beta ^{\prime } &=&2v_{\infty }^{2}\frac{r/r_{opt}^{2}}{\left(
r/r_{opt}\right) ^{2}+r_{0}^{2}}-\frac{4\pi \kappa G}{c^2}\frac{\rho ^{\prime }}{1+\left(8\pi
G\kappa /c^2\right)\rho },\nonumber\\
\end{eqnarray}%
respectively.

Then, by using a series
expansion of $a$ and $\beta ^{\prime }$ to first order of $\kappa $, we
obtain
\be
\left(1+8\pi \frac{G}{c^2}\kappa \rho \right)^{-1/2}+\sqrt{1+8\pi \frac{G}{c^2}\kappa \rho }-2\approx 0,
\ee
\bea
&&2+\sqrt{1+8\pi \frac{G}{c^2}\kappa \rho }-3\left(1+8\pi \frac{G}{c^2} \kappa \rho \right)^{-1/2}\approx
16\pi \frac{G}{c^2}\kappa\rho ,\nonumber\\
\eea
 and
 \be
\beta ^{\prime }=2\frac{v_{\infty }^{2}}{\left( r/r_{opt}\right)
^{2}+r_{0}^{2}}\frac{r}{r_{opt}^{2}}-\frac{4\pi \kappa G}{c^2}\rho ^{\prime }.
\end{equation}
Therefore the field equations Eqs.~(\ref{g1})-(\ref{g2}) become
\be
\frac{d}{dr}\left( re^{-\alpha }\right) =1-8\pi \rho r^{2}, \\ \label{d2}
\ee
and
\bea\label{d3}
&&re^{-\alpha }=\frac{1}{1/r+2v_{\infty }^{2}\left( r/r_{opt}^{2}\right) /%
\left[ \left( r/r_{opt}\right) ^{2}+r_{0}^{2}\right] -4\pi \kappa \rho
^{\prime }},\nonumber\\
\end{eqnarray}%
respectively.

\subsection{The  density profile and the metric of the dark matter halos in EiBI gravity}

In order to obtain a simpler form of Eqs.~(\ref{d2}) and (\ref{d3}) we introduce a set of dimensionless variables $\left(\eta, \theta \right)$ defined as
\be\label{dim}
r=r_{opt}\eta,  \qquad \rho =\frac{c^2}{8\pi Gr_{opt}^2}\theta.
\ee

Hence Eqs.~(\ref{d2}) and (\ref{d3}) can be written in a dimensionless form as
\be\label{fa0}
\frac{d}{d\eta }\left(\eta e^{-\alpha }\right)=1-\theta \eta ^2,
\ee
and
\be\label{fa1}
\eta e^{-\alpha }=\left(\frac{1}{\eta }+2v_{\infty }^2\frac{\eta }{\eta ^2+r_0^2}-k_0\frac{d\theta }{d\eta }\right)^{-1},
\ee
respectively, where we have denoted
\be
k_0=\frac{\kappa }{2r_{opt}^2}.
\ee

By taking the derivative with respect to $\eta $ of Eq.~(\ref{fa1}), with the use of Eq.~(\ref{fa0}) we obtain the following equation describing the behavior of the dark matter density in the EiBi gravity
\bea\label{21}
\theta ''&=& -\kappa _0 \left(\eta ^2 \theta -1\right) \theta '^2
-\frac{ \left[r _0^2+\eta ^2 \left(2
   v_{\infty}^2+1\right)\right]^2}{\kappa _0 \left(\eta
   ^2+r _0^2\right)^2}\theta +
\nonumber\\
&&\frac{2
   \left(\eta ^2 \theta -1\right)  \left(\eta ^2+r _0^2+2
   \eta ^2 v_{\infty}^2\right)}{\eta  \left(\eta ^2+r _0^2\right)}\theta '+\nonumber\\
   &&\frac{6 v_{\infty}^2}{\kappa _0 \left(\eta
   ^2+r _0^2\right)}+\frac{4 \eta ^2 \left(v_{\infty}^2-1\right)
   v_{\infty}^2}{\kappa _0 \left(\eta ^2+r _0^2\right)^2}.
\eea

Equation~(\ref{21}) must be integrated with the initial conditions $\theta (0)=\theta _c$, and $\theta '(0)$, respectively. The second of Eqs.~(\ref{dim}) provides
\be
\theta _c=\frac{8\pi Gr_{opt}^2}{c^2}\rho _c.
\ee
In the following we will fix the dark matter halo parameters so that $r_{opt}=5\;{\rm kpc}=1.5\times 10^{22}$ cm, and $\rho _c=10^{-24}$ g/cm$^3$. This gives $\theta _c=4.19\times 10^{-7}$. For the tangential velocity in the constant velocity region we adopt the value $v_{\infty}=100 $ km/s, while for $r_0$ we take the value $r_0=1$.

\begin{centering}
\begin{figure}
\includegraphics[scale=0.70]{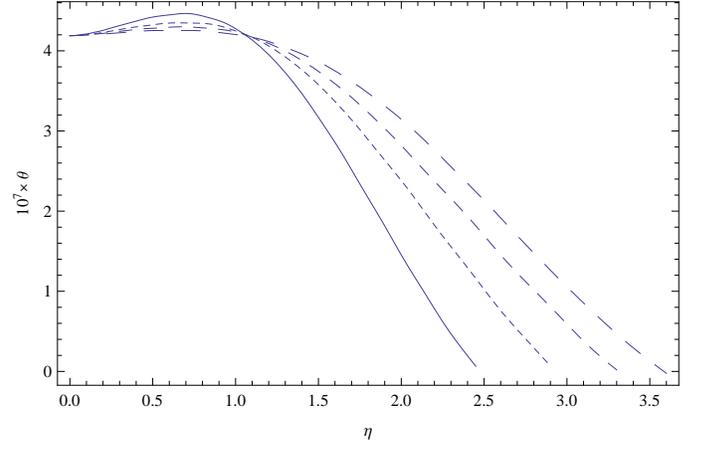}
\caption{Variation of the dimensionless dark matter density profile $\theta $ as a function of the dimensionless radial coordinate $\eta $ for different values of $\kappa _0$: $\kappa _0=0.3 $ (solid curve), $\kappa _0=0.5$ (dashed curve), $\kappa _0=0.7 $ (short dashed curve), and $\kappa _0=1$ (long dashed curve), respectively. The initial conditions are $\theta (0)=4.19\times 10^{-7}$ (corresponding to $r_{opt}=5$ kpc and a dark matter central density $\rho _c=10^{-24}$ g/cm$^3$), and $\theta '(0)=10^{-7}$. For the values of the astrophysical parameters describing the halo we have adopted the numerical values $v_{0\infty}=100$ km/s, and $r_0=1$, respectively.  }\label{fig1}
\end{figure}
\end{centering}

The variation of the dimensionless dark matter density profile $\theta $ as a function of $\eta $, and for different values of $\kappa _0$, is represented in Fig.~\ref{fig1}.
As one can see from Fig.~\ref{fig1}, in the EiBI gravity the dark matter halos have a sharp boundary $\eta _S$, which defines the radius $R_{DM}$ of the galaxy. For the numerical values of the parameter $\kappa _0$ considered in the numerical analysis, we obtain $R_{DM}=\eta _Sr_{opt}\approx 12.5-17.5$ kpc, a value which is reasonable from an astrophysical point of view. Hence, in order to explain the observed properties of the dark matter halos the parameter $\kappa $ of the theory must have a value of the order
\be
\kappa\approx 2r_{opt}^2\kappa _0\approx 2.25\times 10^{44}\;{\rm cm}^2.
\ee

\begin{centering}
\begin{figure}
\includegraphics[scale=0.70]{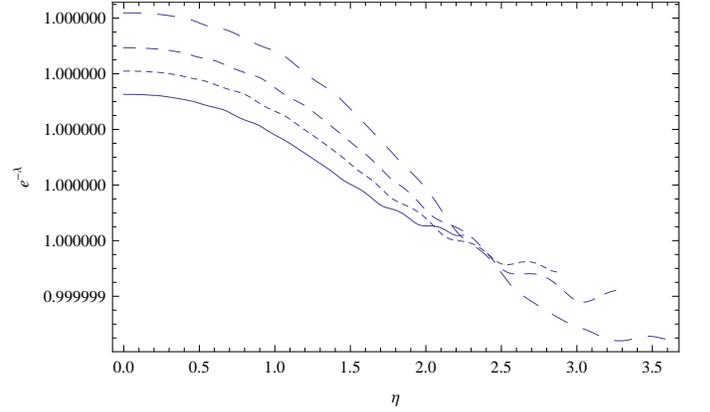}
\caption{Variation of the metric coefficient $e^{-\lambda }=\sqrt{1+2\kappa _0\theta }e^{-\alpha }$ as a function of the dimensionless radial coordinate $\eta $ for different values of $\kappa _0$: $\kappa _0=0.3 $ (solid curve), $\kappa _0=0.5$ (dashed curve), $\kappa _0=0.7 $ (short dashed curve), and $\kappa _0=1$ (long dashed curve), respectively. The initial conditions are $\theta (0)=4.19\times 10^{-7}$ (corresponding to $r_{opt}=5$ kpc and a dark matter central density $\rho _c=10^{-24}$ g/cm$^3$), and $\theta '(0)=10^{-7}$. For the values of the astrophysical parameters describing the halo we have adopted the numerical values $v_{0\infty}=100$ km/s, and $r_0=1$, respectively.  }\label{fig2}
\end{figure}
\end{centering}

The variation of the metric coefficient $e^{-\lambda }=\sqrt{1+2\kappa _0\theta }e^{-\alpha }$ is represented in Fig.~\ref{fig2}.

The mass of the dark matter halo $M_{Dm}(r)$ is given by
\be
\frac{GM_{DM}(r)}{c^2}=\frac{r}{2}\left(1-\sqrt{1+8\pi \frac{G}{c^2}\kappa \rho }e^{-\alpha }\right).
\ee

By introducing the dimensionless dark matter mass $m(\eta)$, defined as
\be
M_{DM}(r)=\frac{c^2r_{opt}}{G}m(\eta ),
\ee
we obtain
\bea
m(\eta )&=&\frac{\eta }{2}\left(1-\sqrt{1+2\kappa _0\theta }e^{-\alpha}\right)=
   \nonumber\\
&&\eta \Bigg(1-\frac{\sqrt{1+2\kappa _0\theta }}{\frac{1}{\eta }+2v_{\infty }^2\frac{\eta }{\eta ^2+r_0^2}-k_0\frac{d\theta }{d\eta }}\Bigg).
\eea

\begin{centering}
\begin{figure}
\includegraphics[scale=0.70]{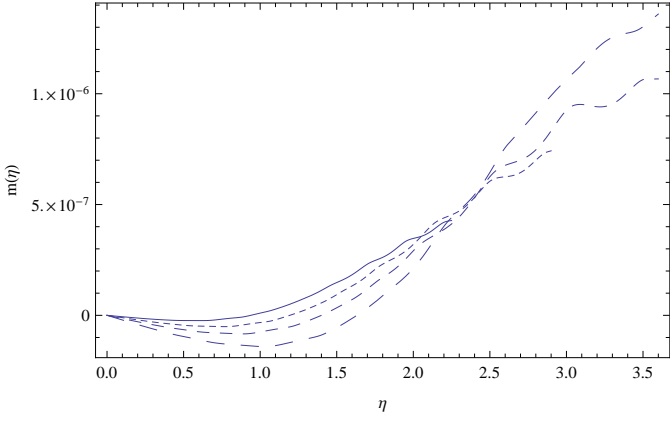}
\caption{Variation of the dimensionless mass profile of the dark matter halo in the EiBI gravity as a function of the dimensionless radial coordinate $\eta $ for different values of $\kappa _0$: $\kappa _0=0.3 $ (solid curve), $\kappa _0=0.5$ (dashed curve), $\kappa _0=0.7 $ (short dashed curve), and $\kappa _0=1$ (long dashed curve), respectively. The initial conditions are $\theta (0)=4.19\times 10^{-7}$ (corresponding to $r_{opt}=5$ kpc and a dark matter central density $\rho _c=10^{-24}$ g/cm$^3$), and $\theta '(0)=10^{-7}$. For the values of the astrophysical parameters describing the halo we have adopted the numerical values $v_{0\infty}=100$ km/s, and $r_0=1$, respectively.  }\label{fig3}
\end{figure}
\end{centering}

The variation of the mass density profile of the dark matter in the EiBI gravity is represented in Fig.~\ref{fig3}.

\section{Dark matter halos in the Newtonian limit}\label{new}

In the non-relativistic limit the action Eq.~(\ref{1c}) leads to the modified Poisson equation for the gravitational potential $\Phi $, given by \cite{1,7},
\be\label{p}
\nabla ^2\Phi=4\pi G\rho +G\frac{\kappa }{4}\nabla ^2 \rho .
\ee

For a spherically symmetric static dark matter distribution, with the use of Eq.~(\ref{p}), we obtain the hydrostatic equilibrium equation as \cite{7}
\be\label{eq}
\frac{dp}{dr}=-\frac{GM}{r^2}\rho-G\frac{\kappa }{4}\rho \rho '.
\ee
In the case of dark matter halos the approximation $p=0$ gives a very good description of the galactic properties of the dark matter. By combining Eq.~(\ref{eq}) with $p=0$ with the mass continuity equation
\be
\frac{dM}{dr}=4\pi \rho r^2,
\ee
we obtain the following equation describing the density distribution of the dark matter halos in the Newtonian approximation of EiBI gravity,
\be
\frac{1}{r^{2}}\frac{d}{dr}\left( r^{2}\frac{d\rho ^{(0)}}{dr}\right) +\frac{%
2}{\bar{\kappa} }\rho ^{(0)}=0, \label{p1}
\ee
where we have denoted by $\rho ^{(0)}$ the density in the Newtonian approximation, and $\bar{\kappa}=\kappa /8\pi $, respectively.

Eq.~(\ref{p1}) is the Lane-Emden equation for the polytropic index $n=1$ %
\cite{pola,polb}, and its non-singular solution at the center is given by
\begin{equation}
\rho ^{(0)}(r)=K\frac{\sin \left( \sqrt{2/\bar{\kappa} }r\right) }{\sqrt{2/\bar{\kappa} }%
r},
\end{equation}%
where $K$ is an arbitrary constant of integration. Since at the center of
the galactic halo the density of the dark matter is $\rho ^{(0)}(0)=\rho
_{0} $, it follows that $K=\rho _{0}$. In the EiBI theory the dark halo has
a sharp boundary $R_{DM}$, corresponding to $\rho ^{(0)}(R_{DM})=0$, which
gives $R_{DM}=\sqrt{\bar{\kappa} /2}\pi $. Thus, the mass profile of the dark
matter $M(r)=4\pi \int_{0}^{r}{\rho ^{(0)}(r)r^{2}dr}$ is given by
\begin{equation}
M(\bar{r})=\frac{4R_{DM}^{3}}{\pi ^{2}}\rho _{0}\left[ \sin \left( {\bar{r}}%
\right) -{\bar{r}}\cos \left( {\bar{r}}\right) \right] ,
\end{equation}%
where ${\bar{r}}=\pi r/R_{DM}$ is defined for convenience.

The $q$-metric coefficient $e^{-\alpha }=1-2GM(r)/c^2r$ yields
\bea
e^{-\alpha }&=&1-\frac{8R_{DM}^{2}}{\pi {\bar{r}}}\rho _{0}\left[ \sin \left( {%
\bar{r}}\right) -{\bar{r}}\cos \left( {\bar{r}}\right) \right]=
    \nonumber\\
&&1-\frac{{%
\bar{\rho}}_{0}}{{\bar{r}}}\left[ \sin \left( {\bar{r}}\right) -{\bar{r}}%
\cos \left( {\bar{r}}\right) \right] ,
\eea
where ${\bar{\rho}}_{0}=8G\rho _{0}R_{DM}^{2}/\pi c^2$.
For the metric coefficient $e^{-\lambda }=e^{-\alpha }a$, we obtain
\begin{equation}
e^{-\lambda }\approx e^{-\alpha }\left( 1+4\pi \kappa \frac{G}{c^2}\rho \right)
=e^{-\alpha }\left[ 1+\frac{{\bar{\rho}}_{0}}{{\bar{r}}}\sin \left( {\bar{r}}%
\right) \right] ,
\end{equation}%
 while the function $f(\bar{r})$ is given by
\begin{equation}
f(\bar{r})=\frac{\left( R_{DM}^{2}/\pi ^{2}\right) \bar{r}^{2}}{\sqrt{1+8\pi
\kappa \rho }}\approx \frac{R_{DM}^{2}}{\pi ^{2}}\bar{r}^{2}\left[ 1-\frac{{%
\bar{\rho}}_{0}}{{\bar{r}}}\sin \left( {\bar{r}}\right) \right] .
\end{equation}

Thus, the metric of the dark matter halo takes the form %
\begin{eqnarray}
&&ds^{2} =-e^{\nu _{0}}\left[ \left( \frac{R_{DM}}{\pi r_{opt}}\right) ^{2}%
\bar{r}^{2}+r_{0}^{2}\right] ^{v_{\infty }^{2}}c^2dt^{2}  +\notag \\
&&\frac{R_{DM}^{2}}{\pi ^{2}}\frac{d\bar{r}^{2}}{\left[ 1-\frac{{\bar{\rho}}%
_{0}}{{\bar{r}}}\sin \left( {\bar{r}}\right) +{\bar{\rho}}_{0}\cos \left( {%
\bar{r}}\right) \right] \left[ 1-\frac{{\bar{\rho}}_{0}}{{\bar{r}}}\sin
\left( {\bar{r}}\right) \right] }  +\notag \\
&&\frac{R_{DM}^{2}}{\pi ^{2}}\bar{r}^{2}\left[ 1-\frac{{\bar{\rho}}_{0}}{{%
\bar{r}}}\sin \left( {\bar{r}}\right) \right] d\Omega ^{2}.
\end{eqnarray}

The total mass of the dark matter halo is given by $M_{DM}=M\left(R_{DM}%
\right)=\sqrt{2} \pi ^2 \kappa ^{3/2} \rho _0=\left(4/\pi\right)\rho
_0R_{DM}^3 $. The mean density $<\rho >$ of the dark matter halo is obtained
as $<\rho >=3M_{DM}/4\pi R_{DM}^3=3\rho _0/\pi^2$.

In the Newtonian approximation the gravitational potential $V_{grav}\left(
r\right) $ of the dark matter distribution in the EiBI theory is determined for $\bar{r}/\pi \leq 1$ by
\begin{equation}
V_{grav}=\frac{1}{8\pi }\int_{r}^{R_{DM}}\frac{M\left( r^{\prime }\right)
dr^{\prime }}{r^{\prime 2}}=\frac{{\bar{\rho}}_{0}}{16\pi {\bar{r}}}\sin
\left( {\bar{r}}\right) .
\end{equation}
 At small radii and for $\bar{r}/\pi \leq 1$  the potential behaves as
\begin{equation}
V_{grav}({\bar{r}})\approx \frac{{\bar{\rho}}_{0}}{16\pi }\left( 1-\frac{1}{%
3!}{\bar{r}}^{2}+\frac{1}{5!}{\bar{r}}^{4}-\frac{1}{7!}{\bar{r}}^{6}\right)
+O({\bar{r}})^{8}.
\end{equation}

The gravitational potential energy $U(r)$ per unit mass and inside radius $r$
of the dark matter halo yields
\begin{eqnarray}
U &=&-\frac{1}{2}\int_{0}^{r}\frac{\rho ^{(0)}(r)M(r)}{r}r^{2}dr=-\frac{%
R_{DM}^{5}\rho _{0}^{2}}{4\pi ^{3}}\times  \notag \\
&& \left\{ 2\bar{r}\left[ 2+\cos \left( 2\bar{r}\right) \right] -3\sin
\left(2\bar{r}\right) \right\} .
\end{eqnarray}
Thus, the total potential energy of the dark matter halo is given by $%
U\left( R_{DM}\right) =-\left( 3\rho _{0}^{2}/2\pi ^{2}\right) R_{DM}^{5}$.

\section{Discussions and final remarks}\label{sect5}

In the present paper, we have analyzed the dark matter
density profiles, and their geometry, in the EiBI gravitational theory. As a first step in our study we have adopted a very general tangential velocity profile \cite{per}, which, together with the requirement of the motion of the test particles in stable circular orbits, allows to formulate the basic equations describing the general relativistic dark matter halos in the EiBI theory. The solution of the field equations has been obtained numerically, after reducing them to a dimensionless form.

In the Newtonian approximation of the dark matter pressure satisfying the condition $p=0$,  the dark matter density profile satisfies the Lane-Emden equation with a polytropic index $n=1$ \cite{pola,polb}. Therefore the dark matter profile can be obtained in an exact analytic form, as well as the full galactic dark matter metric, in both $g$ and $q$ geometries. Similar results for pressureless stellar profiles have been obtained in \cite{Pani}.
The non-singular $n=1$ polytropic density profiles show the presence of an extended core, whose presence in the considered model is due to the matter-gravity coupling, and modified geometry.

In fact, in the context of extended theories of gravity, it was argued that the generic corrections to the Newtonian potential could fit data and reproduce the observational galactic dynamics \cite{Capozziello:2012ie}. These corrections are not phenomenological but arise from the weak field limit of the extended relativistic theories of gravity that predict the existence of Yukawa-like corrections to the Newtonian potential. These corrections imply that further scales have to be taken into account and that their effects should be irrelevant at local Solar System scales. Indeed, considering the modified Lane-Emden equation arising from $f(R)$ gravity, it was shown that the differences between GR and $f(R)$ gravitational potentials become evident for larger radii from the centre, so that the corrections to the potential can significantly boost the circular velocity for an extended system such as in a galaxy. In the strong field regime, such corrections could give rise to peculiar stellar structure or trigger the Jeans instability, and this issue was further explored in \cite{Capozziello:2011gm}.

As follows from our numerical results, in order to obtain realistic dark matter density profiles the unique free parameter of the EiBI gravity must be of the order of $\kappa \approx 10^{44}$ cm$^2$. This result is consistent with the value of the galactic halo radius obtained in the Newtonian approximation. On the other hand the studies of the structure of the compact general relativistic stars require a value of $\kappa $ of the order of $\kappa =10^{12}$ cm$^2$ \cite{new, 8na,8nb,8nc}. Moreover, for a given $\kappa $, the theory predicts a universal dark matter halo distribution. These contradictory results raise the question of the viability of the EiBI gravity as a theoretical model that could describe gravitational phenomena on all astrophysical and cosmological scales.  It is interesting to note that similar constraints requiring very large values of $\kappa $ have been obtained in \cite{3} from the requirement of the absence of strong near-surface curvature effects. For
simple Newtonian configurations in which the density distribution
deviates from perfect smoothness at a given cut--off surface the EiBI theory may yield surface singularities. For the case of a polytropic star with $M=1.4M_{\odot}$ and radius $R=10^{-2}R_{\odot}$, the constraint on $\kappa $ can be formulated as $\kappa \geq 6\times 10^{38+2k}$ cm$^2$,   where $k$ is a parameter related to the form of the polytropic equation of state that provides a reliable description of the matter close to the surface of the star.

In the standard dark matter models, the $n=1$ polytrope is used to model
dark matter in the form of a Bose-Einstein condensate, namely, an assembly
of light individual bosons that acquire a repulsive interaction by occupying
the same ground energy state \cite{Bosea,Boseb,Bosec}.

 In the present approach to the dark matter halos all the relevant astrophysical
quantities can be predicted from the model, and can be directly compared
with the corresponding observational parameters (the dark halo mass, the
radius of the galaxy, etc), with all the properties of the dark matter halos
being determined by a single parameter $\kappa $. Therefore, although
lying beyond the range of the present paper, the in-depth comparison of the theoretical
predictions of the EiBI gravity model with the galactic and extra galactic
scale observations yield tight constraints on the numerical values of $%
\kappa $, thus leading to a direct viability test of the theory. Work along
these lines is currently underway.

\section*{Acknowledgments}

FSNL is supported by a Funda\c{c}\~{a}o para a Ci\^{e}ncia e Tecnologia
Investigador FCT Research contract, with reference IF/00859/2012, funded by
FCT/MCTES (Portugal).
FSNL also acknowledges financial support of the Funda%
\c{c}\~{a}o para a Ci\^{e}ncia e Tecnologia through the grants
CERN/FP/123615/2011 and CERN/FP/123618/2011. SVS acknowledges financial
support of the Russian Foundation for Basic Research through grants No.
11-02- 01162 and 13-02-12093 .

\end{document}